\documentclass[%
reprint,
showpacs,
 amsmath,amssymb,
 aps,
 prl,
 10pt,
 twocolumn,
 floats,
]{revtex4-1}

\usepackage[latin1]{inputenc}
\usepackage{graphicx}
\usepackage{bm}
\usepackage{color}

\newcommand{\ket}[1]{\ensuremath{ \left|#1\right\rangle}}

\begin{document}


\title{Fiber ring resonator with nanofiber section for \\ chiral cavity quantum electrodynamics and multimode strong coupling}

\author{Philipp Schneeweiss}
\email{schneeweiss@ati.ac.at}
\author{Sophie Zeiger}
\author{Thomas Hoinkes}
\author{Arno Rauschenbeutel}
\author{J\"urgen Volz}
\email{jvolz@ati.ac.at}
\affiliation{%
 \mbox{Vienna Center for Quantum Science and Technology, Atominstitut, TU Wien, Stadionallee 2, 1020 Vienna, Austria}
}%

\date{\today}

\begin{abstract}
We experimentally realize an optical fiber ring resonator that includes a tapered section with subwavelength-diameter waist. In this section, the guided light exhibits a significant evanescent field which allows for efficient interfacing with optical emitters. A commercial tunable fiber beam splitter provides simple and robust coupling to the resonator. Key parameters of the resonator such as its out-coupling rate, free spectral range, and birefringence can be adjusted. Thanks to the low taper- and coupling-losses, the resonator exhibits an unloaded finesse of $F=75\pm1$, sufficient for reaching the regime of strong coupling for emitters placed in the evanescent field. The system is ideally suited for trapping ensembles of laser-cooled atoms along the nanofiber section. Based on measured parameters, we estimate that the system can serve as a platform for optical multimode strong coupling experiments. Finally, we discuss the possibilities of using the resonator for applications based on chiral quantum optics.

\end{abstract}
\maketitle



Over the past years, significant research effort has been devoted to interfacing quantum emitters, such as molecules, quantum dots, color centers, and neutral atoms, with fiber-guided light fields. Suitable light--matter interfaces are considered to be key elements for future quantum networks~\cite{Kimble08}. One way to realize such an interface consists in coupling emitters to the evanescent fields surrounding the nanofiber-waist of a tapered optical fiber, i.e., a fiber section with sub-wavelength diameter. Such systems already provide absorption probabilities for single fiber-guided photons of about 25~\% for a single emitter located on the nanofiber surface~\cite{Yalla12} and about 5~\% at a distance of 200~nm, typical for cold atoms in nanofiber-based optical dipole traps~\cite{Vetsch10,Goban12}.

One way to further enhance the light--matter coupling strength is to increase the number of emitters in the evanescent field and to take advantage of their collective coupling. Another option relies on confining the light in an optical resonator, which allows one to even reach the regime of strong coupling in the sense of cavity quantum electrodynamics (CQED)~\cite{Berman1994}. There, coherent emitter-light interaction strength dominates over the incoherent decay channels. In this context, optical nanofibers are a versatile platform as they allow one to combine both approaches and, in this way, to reach very strong light--matter coupling in an fiber-integrated environment.

Different nanofiber-based Fabry-Pérot resonator schemes have been developed, e.g., based on Bragg structures created using ion beam milling~\cite{Ding11} or laser ablation~\cite{Nayak14} of the fiber waist. A nanofiber placed on an optical grating has been demonstrated and Purcell enhancement of single quantum emitters coupled to this system has been observed~\cite{Yalla14}. Tapered optical fibers have been combined with conventional fiber Bragg gratings to form a Fabry-Pérot resonator~\cite{Wuttke12} for which strong coupling has been demonstrated with Cesium (Cs) atoms trapped close to the nanofiber surface~\cite{Kato15}. Running-wave type resonators such as nanofiber knot and loop resonators~\cite{Brambilla10} as well as a nanofiber-segment tapered fiber closed to a ring via a 50:50 beam splitter~\cite{Jones16} have also been demonstrated, albeit so far with smaller finesse than their Fabry-Pérot counterparts.

Here, we demonstrate a tapered fiber-based ring resonator with optical characteristics that are compatible with entering the regime of single-atom strong coupling. We experimentally reach a resonator finesse of $F=75\pm1$ which corresponds to a single-atom cooperativity of $C\approx1$. Our implementation offers easy tuning of the resonator eigenpolarizations and out-coupling rate as well as straightforward adjustment of the resonator's free spectral range. Our system is compatible with established nanofiber-based schemes for the optical trapping of laser-cooled atoms. This would enable a controlled coupling of large ensembles of atoms to the resonator field which then gives rise to extremely large collective atom-resonator interaction strengths. Furthermore, the evanescent fields around the nanofiber exhibit an inherent link between the local polarization and the propagation direction of the guided light~\cite{LeKien14a}. This gives rise to a direction-dependent light--emitter coupling~\cite{Mitsch14b,Petersen14} which renders this resonator distinct from traditional Fabry-P\'erot or ring-resonators. In particular, this allows one to implement chiral quantum optics effects~\cite{Arxiv_Lodahl16}. Finally, the optical path length of such a tapered fiber ring (TFR) resonator can be significantly increased without reducing its cooperativity. This makes this system a prime candidate for experimentally exploring the regime of optical multimode strong coupling where the atoms simultaneously couple strongly to many longitudinal resonator modes~\cite{Parker87,Krimer14}.  

\begin{figure}[tb]%
\includegraphics[width=1.0\columnwidth]{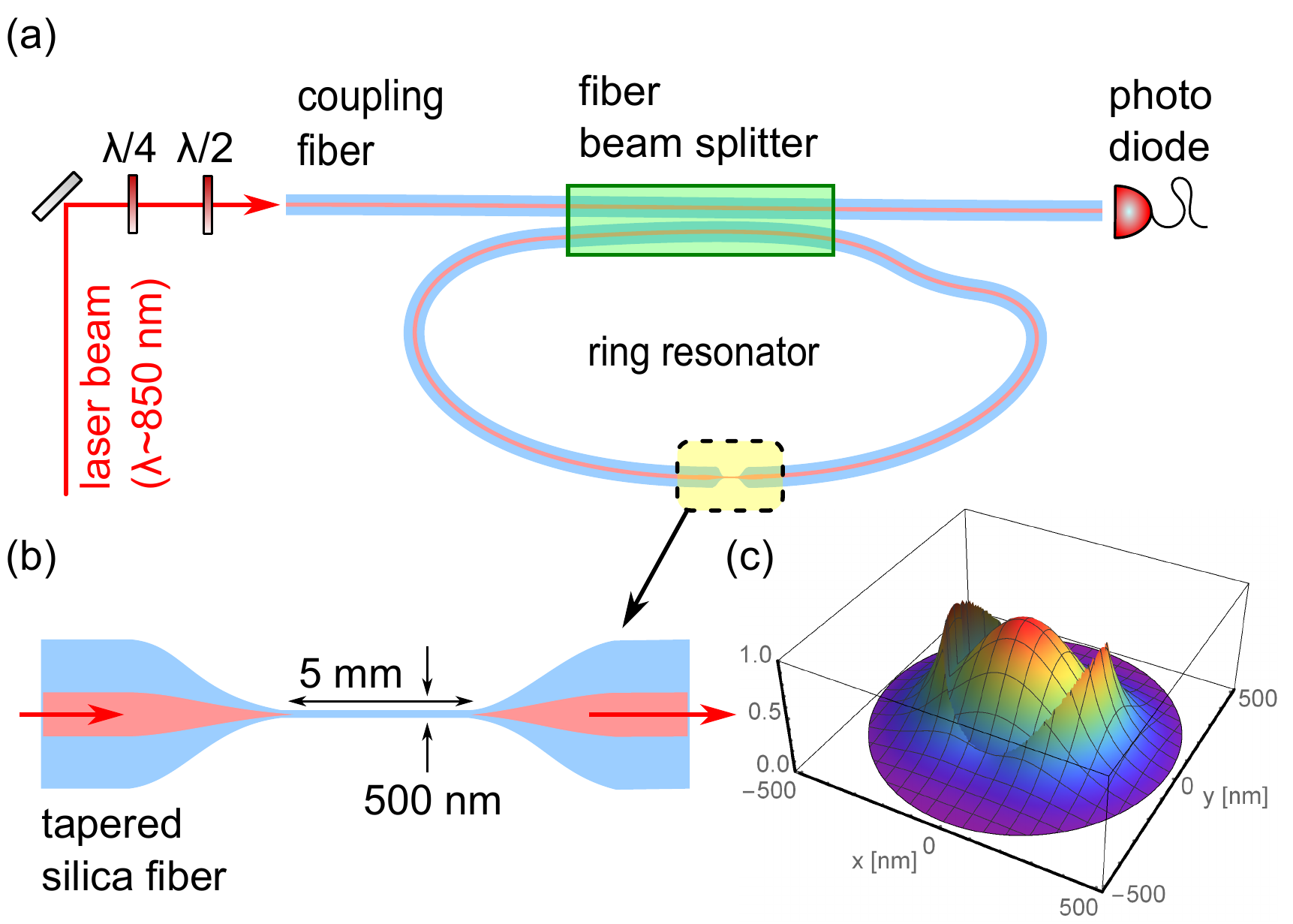}%
\caption{(a) Sketch of the experimental setup: A tapered optical fiber with a nanofiber waist forms a fiber ring resonator. An adjustable fiber beam splitter allows one to couple light into and out of the resonator. The experimental characterization of the system is carried out by recording transmission spectra through the coupling fiber around a center wavelength of about 850~nm. The polarization of the input light field can be adjusted using a half and a quarter wave plate. (b) Schematic of the tapered optical fiber section with a nanofiber waist. A waist with a diameter of $500$~nm and a length of $5$~mm was used in the experiment. (c) Normalized intensity profile of a quasi-linearly polarized nanofiber-guided light field. The pronounced evanescent part of the guided light allows efficient interfacing with optical emitters. (parameters: nanofiber radius 250~nm; vacuum optical wavelength 852~nm; fiber refractive index 1.45.)
}%
\label{Fig:Setup}%
\end{figure}

Our ring resonator consists of a tapered optical fiber with a nanofiber waist connected to an adjustable fiber beam splitter (Newport F-CPL-830-N-FA), see Fig.~\ref{Fig:Setup}(a), whose splitting ratio can be continuously adjusted between 0~\% and 100~\%. The tapered fiber, shown schematically in Fig.~\ref{Fig:Setup}(b), is fabricated from a standard step-index silica fiber (Fiber~Core SM~800) using a heat-and-pull process~\cite{Birks92,Warken08}. The fiber waist has a length of 5~mm and a diameter of 500~nm. Light guided in such a subwavelength-diameter optical fiber exhibits a pronounced evanescent field (Fig.~\ref{Fig:Setup}(c)) and enables efficient, homogeneous coupling of optical emitters to the guided mode. The standard fiber ends of the tapered fiber are spliced to the fiber beam splitter, yielding a TFR resonator with a total resonator length of $2.35\pm0.01$ m. The remaining two outputs of the fiber beam splitter then constitute the coupling fiber that allows us to interface the resonator.

\begin{figure}[t]%
\includegraphics[width=1\columnwidth]{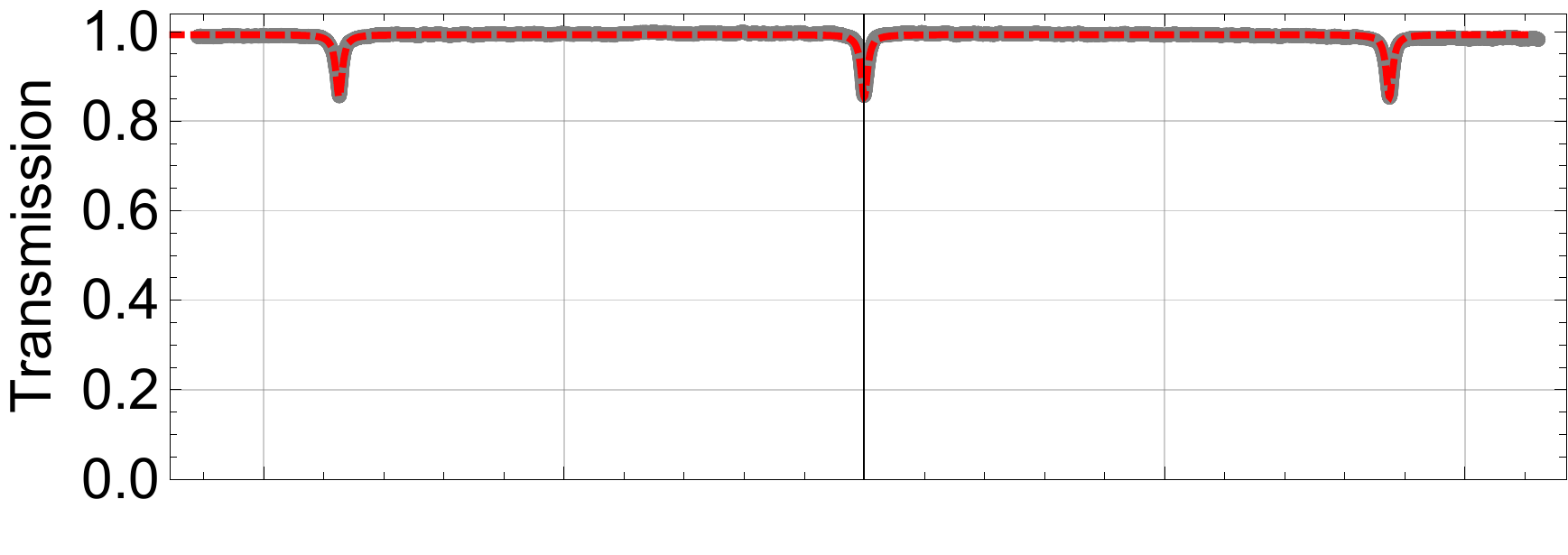}\\ \vspace{-0.2cm}
\includegraphics[width=1\columnwidth]{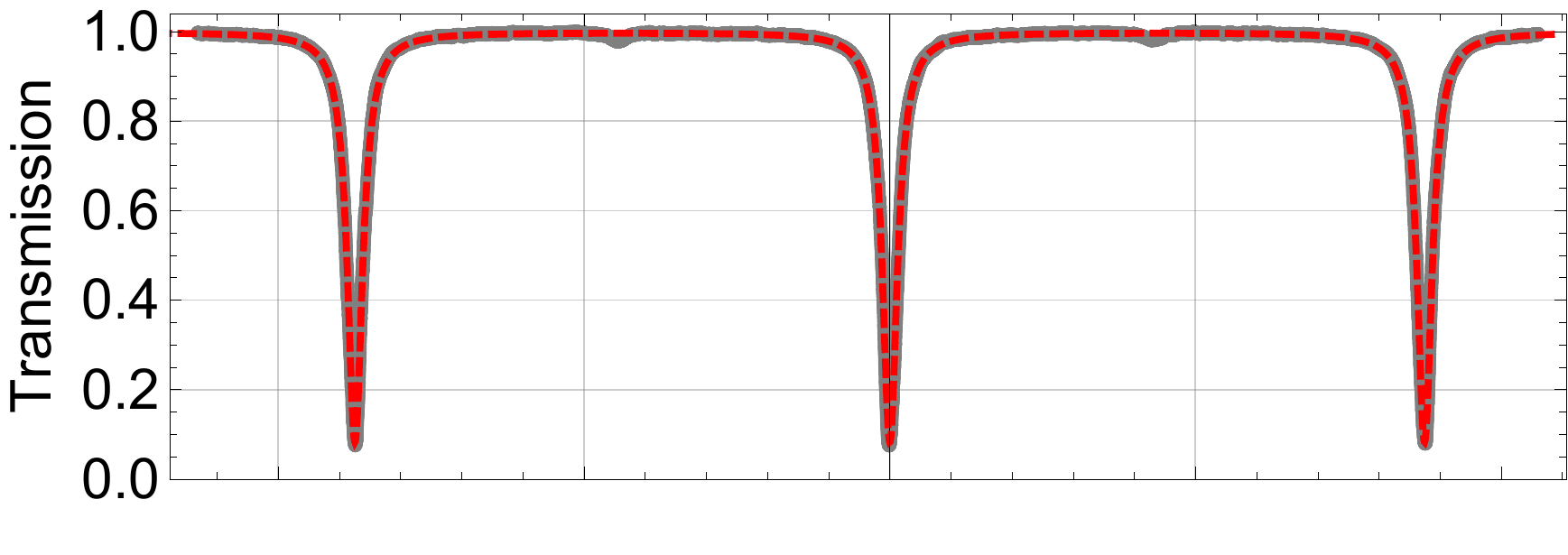}\\ \vspace{-0.2cm}
\includegraphics[width=1\columnwidth]{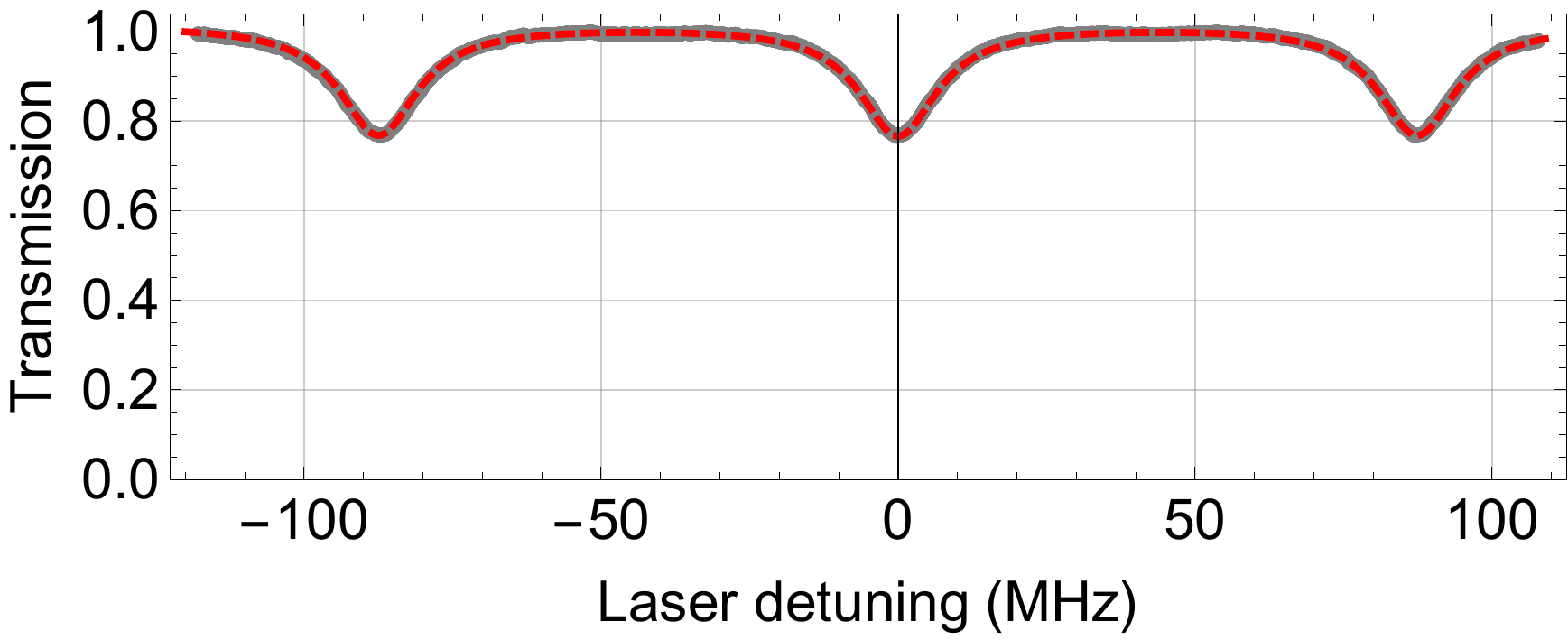}
\caption{%
Measured coupling fiber transmission (black) as a function of the laser detuning for the (a) undercoupled regime ($\kappa_{\rm ext}<\kappa_0$), (b) critically coupling ($\kappa_{\rm ext}=\kappa_0$), and (c) overcoupled regime ($\kappa_{\rm ext}>\kappa_0$). The red dashed lines are Lorentzian fits to the data.
}%
\label{Fig:Spectra}%
\end{figure}

We optically characterize the TFR resonator by recording transmission spectra through the coupling fiber. To this end, we send light from a tunable, narrow-band diode laser into the coupling fiber and align the probe light's polarization with one of the principle polarization axes of the resonator using wave plates. The transmitted power is recorded with a photodiode as a function of the laser--resonator detuning. The wavelength of the light is about $850$~nm, close to the Cs D2 line. By adjusting the splitting ratio of the beam splitter, we can  continuously adjust the fiber--resonator coupling rate $\kappa_{\rm ext}$ from the undercoupled regime ($\kappa_{\rm ext}<\kappa_0$) to critical coupling ($\kappa_{\rm ext}=\kappa_0$) and to the overcoupled regime ($\kappa_{\rm ext}>\kappa_0$). Here, $\kappa_0$ is the unloaded field decay rate of the uncoupled TFR resonator. 
Figure~\ref{Fig:Spectra} shows example spectra for the three different coupling regimes. In each case, the laser is scanned over three consecutive resonances. Narrow transmission dips are clearly apparent with a free spectral range (FSR) of  $\nu_\textrm{FSR}=87.5\pm0.4$~MHz demonstrating a high optical finesse and indicating a small internal resonator field decay rate $\kappa_0$. We checked that the birefringence of the resonator and, thus, its eigenpolarizations can be tuned using a fiber paddle polarization controller, and coupling to only one of the resonator modes can always be restored by adapting the setting of the wave plates in the coupling beam path.

\begin{figure}[t]%
\includegraphics[width=1\columnwidth]{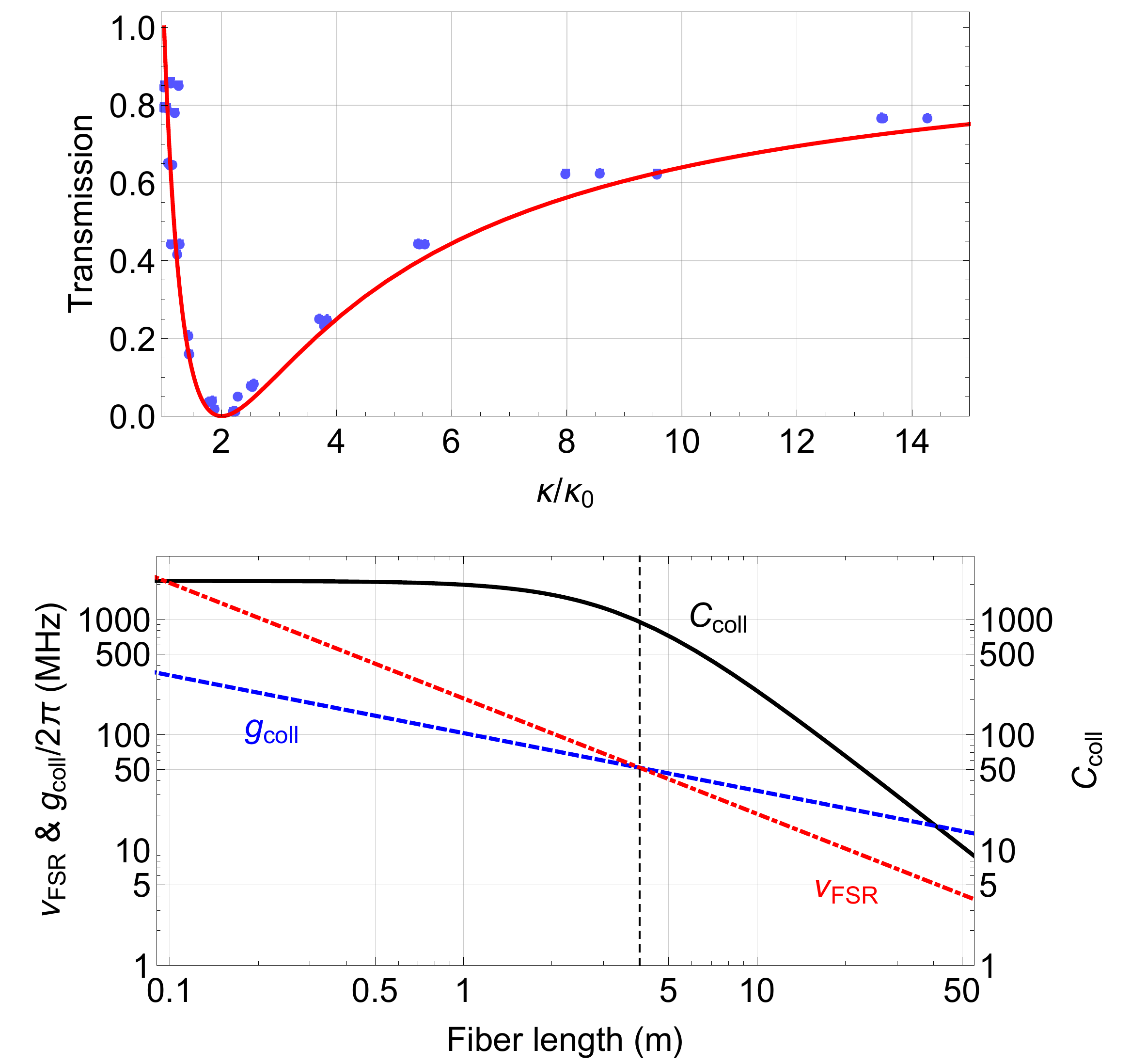}%
\caption{ (a) Transmission through the coupling fiber as a function of the total field decay rate of the resonator $\kappa=\kappa_{\rm ext}+\kappa_0$. From small to large values of $\kappa$, the regimes of undercoupling, critical coupling, and overcoupling are observed. The red dashed curve is a fit using Eq.~\ref{Eq:t}. Error bars, indicating the standard deviation of the $\{\kappa, T_r \}$ fit values, are small and hardly visible. (b) Expected collective coupling strength $g_\textrm{coll}$ (blue dashed line), free spectral range $\nu_\textrm{FSR}$ (red dash-dotted line) and cooperativity $C_\textrm{coll}$ (black solid line) of the TFR resonator as a function of the resonator length. The plot is calculated for fiber losses of 1.5 dB/km for optical fibers at a wavelength of about 850~nm~\cite{Goff13} and for the case of a state-of-the-art nanofiber trap loaded with 2000 atoms~\cite{Vetsch10}. The dashed vertical line indicates the length at which the $g_\textrm{coll}$ exceeds the free spectral range, and the atom-resonator system enters the regime of multimode strong coupling.
}%
\label{Fig:ToverKappa}%
\end{figure}

Close to a cavity resonance, the field transmission $t$ through the coupling fiber as a function of the frequency of the probe light is given by the Lorentzian~\cite{Haus84}, 
\begin{align}
t&=\frac{\kappa_0 - \kappa_{\rm ext} + i(\omega-\omega_0)}{\kappa_0 + \kappa_{\rm ext} + i(\omega-\omega_0)}\;,
\label{Eq:t}
\end{align}
and is related to the power transmission $T$ by $T=|t|^2$. Here, $(\omega-\omega_0)$ is the detuning between the laser frequency $\omega$ and the resonator frequency $\omega_0$. In order to precisely determine the unloaded resonator loss rate, we record transmission spectra for a total of 12 different settings of the fiber beam splitter. For each setting, the laser is scanned over three consecutive resonances. All spectra and all resonances are then fitted with the Lorentzian in Eq.~(\ref{Eq:t}) from which we obtain the overall field decay rate $\kappa=\kappa_0 + \kappa_{\rm ext}$ and the on-resonance transmission $T_r$.  This dataset, $\{\kappa, T_r \}$, is shown as blue dots in Fig.~\ref{Fig:ToverKappa}.
By adjusting the splitting ratio of the fiber beam splitter and, thus, increasing $\kappa$, we clearly observe the transition from undercoupling to critical coupling and to overcoupling. In order to infer the unloaded field decay rate of the resonator $\kappa_0$, we fit the theoretical prediction in Eq.~\ref{Eq:t} to the data in Fig.~\ref{Fig:ToverKappa} with $\kappa_0$ as the only free parameter. The fit shows excellent agreement with the data which demonstrates that the fiber beam splitter does not introduce significant parasitic losses. From the fit, we obtain $\kappa_0=2\pi\times 0.58 \pm 0.01$~MHz which yields an unloaded finesse of $F=\pi\nu_\textrm{FSR}/\kappa_0 =75\pm 1$ and a quality factor of about $Q=3\times10^8$. 

Based on the experimental characterization of the resonator, we now estimate its potential for the realization of a light--matter interface and for experiments in the realm of CQED. We focus on the typical example of interfacing the fiber with an ensemble of $^{133}$Cs atoms coupled to the evanescent field of the resonator modes~\cite{Vetsch10}. The closed optical transition $\ket{F = 4,m_{F}=4} \leftrightarrow \ket{F^\prime = 5,m_{F^\prime}=5}$  from the $6S_{1/2}$ ground to the $6P_{3/2}$ excited state (wavelength $\lambda=852$ nm, dipole decay rate $\gamma=2\pi\times2.6$ MHz) is $\sigma^+$-polarized. For a suitably chosen quantization axis, this optical transition has almost unit polarization overlap with the nanofiber-guided light~\cite{Mitsch14b}. For this setting, we find an emitter--light coupling strength $g=2\pi\times1.5$~MHz for an atom located 200~nm away from the nanofiber surface, i.e., at a typical atom--surface separation realized with nanofiber-based traps~\cite{Vetsch10}. For an atom at the surface we find a coupling strength of $g_\textrm{surf}=2\pi\times 5$~MHz. The former coupling strength corresponds to a single-atom cooperativity $C_0=g^2/(2\kappa\gamma)=0.74$, close to the regime of strong coupling. Employing established techniques, it should be feasible to trap about 2000 atoms close to the nanofiber~\cite{Vetsch10}. This would realize a large collective interaction between the atomic ensemble and the light in the resonator, yielding a collective cooperativity of $C_{\rm coll}=NC_0\approx 1500$ which sets the system far in the strong-coupling regime.

We now estimate how the resonator and coupling parameters change with the resonator length. Diffraction effects make it challenging to maintain a small mode area (and, thus, large coupling strength) for long resonators build from free-space optical components. However, for both, Fabry-Pérot and ring-type fiber-based resonators, the light is \emph{guided} and the mode cross section is independent of the resonator length. Then, the atom-resonator coupling strength scales as $g\propto l^{-1/2}$ with the length $l$ of the resonator. At the same time, the cumulative propagation losses in the standard fiber part are typically much smaller than the losses that occur when the light propagates through the tapered section. In this case, the unloaded resonator decay rate $\kappa_0$ depends on the length as $\kappa_0\propto l^{-1}$. Consequently, the cooperativity is then independent of the resonator length. Remarkably, it is thus possible to maintain large cooperativities also for large resonator lengths. For our experimental system, Fig.~\ref{Fig:ToverKappa} shows the predicted behavior of the collective cooperativity $C_\textrm{coll}$ as well as collective coupling strength $g_\textrm{coll}=\sqrt{N}g$ and free spectral range $\nu_\textrm{FSR}$ as a function of the geometrical resonator length of the TFR resonator. The calculation takes fiber propagation losses into account. For fiber lengths of about $l>4$~m, the collective coupling strength $g_\textrm{coll}$ exceeds the free spectral range and the system enters the regime of multimode strong coupling~\cite{Parker87}. In this regime of CQED, the atomic ensemble is simultaneously strongly coupled to several non-degenerate longitudinal resonator modes. So far, this has only been realized in the microwave domain~\cite{Sundaresan15}. Multimode strong coupling enables atom-mediated interactions between different optical modes and results in a nonlinear quantum dynamics that is not present in the single-mode regime~\cite{Krimer14}. Atom-mediated mode coupling has recently also been considered for implementing photonic quantum simulation protocols~\cite{Schine2016}.

Apart from the large possible collective coupling strength and the intrinsic fiber integration, a TFR resonator differs from other resonator types because of the extremely tight confinement of the light in the nanofiber section. This confinement gives rise to an inherent link between local polarization and propagation direction of the light. In conjunction with suitable quantum emitters, such as spin-polarized atoms, NV-centers or quantum dots, this gives rise to strong direction-dependent light--matter interaction strengths~\cite{Mitsch14b,Petersen14}, similar to the case of whispering-gallery-mode resonators~\cite{Junge13}. This renders the TFR resonator a conceptually novel type of optical resonator for which collectively enhanced chiral light-matter interaction~\cite{Arxiv_Lodahl16} provides novel quantum functionalities~\cite{Shomroni14} and that can, e.g., be employed to realize optical nonreciprocal devices~\cite{Sayrin15b,Scheucher2016}. 

In summary, we realized a tapered fiber ring resonator that incorporates a nanofiber section, and identified two key areas of application. The high optical finesse, the tight confinement of the light in the nanofiber section, and the compatibility with homogeneously coupling to large ensembles of optical emitters render this system well suited for protocols that require strong light--matter interaction. Our experimental approach also offers practical advantages such as a simple tuning of key system parameters. Both, the chiral light--matter interactions present in our resonator and the option to enter the regime of multimode strong coupling, make tapered fiber ring resonators a powerful platform for future experiments in classical and quantum photonics.

We acknowledge financial support by the Austrian Science Fund (FWF, SFB NextLite project No. F~4908-N23).


\bibliography{Literature}

\end{document}